\documentclass[preprint,aps,showpacs,nofootinbib,preprintnumbers,amsmath,amssymb]{revtex4-1}
\usepackage{epsfig}
\usepackage{bm}
\usepackage[usenames ,dvipsnames]{xcolor}
\usepackage{slashed}
\usepackage{caption}
\usepackage{subcaption}
\usepackage{graphicx,color}

\usepackage{multirow}
\usepackage{tabularx}

\begin{document}
	
	\title{Study of  charmless two-body baryonic $B$ decays}

\author{Xiang-Nan Jin$^{1,2}$\footnote{jinxiangnan21@mails.ucas.ac.cn},  Chia-Wei Liu$^{2}$\footnote{chiaweiliu@ucas.ac.cn}
 and Chao-Qiang Geng$^{2}$\footnote{cqgeng@ucas.ac.cn}}
\affiliation{
	$^{1}$International Centre for Theoretical Physics Asia-Pacific (ICTP-AP), UCAS, Beijing 100190, China \\
	$^{2}$School of Fundamental Physics and Mathematical Sciences, Hangzhou Institute for Advanced Study, UCAS, Hangzhou 310024, China 
}
\date{\today}

\begin{abstract}
We  study the charmless two-body decays of  $B\to {\bf B}_n \overline{\bf B}_n^\prime$ with $B = ( B^+ , B^0 , B^0_s)$ and ${\bf B}_n^{(\prime)}$ the low-lying octet baryons. The factorizable amplitudes are calculated by the modified bag model, while the nonfactorizable ones are extracted from the experimental data with the $SU(3)_F$ flavor symmetry. We are able to explain all current experimental measurements and provide predictions on the decay branching ratios in $B\to {\bf B}_n \overline{\bf B}_n^\prime$. Particularly, we find that ${\cal B}(B_s^0 \to \Xi^0 \overline{\Xi}^0) = (32.5\pm 3.7 )\times 10^{-7}$ and ${\cal B}(B_s^0 \to \Xi^- \overline{\Xi}^-) = (33.1\pm 3.8 )\times 10^{-7}$, which are sizable and able to be observed in the ongoing experiments at LHCb and BELLE-II. Furthermore, the decay branching ratios of $B_s^0 \to (p\overline{p}, n\overline{n})$ and $B^0 \to (\Xi^0 \overline{\Xi}^0 ,\Sigma^+\overline{\Sigma}^+)$ are expected to be $O(10^{-10})$, which are suppressed  due to the angular momentum conservation and  chiral symmetry.
\end{abstract}

\maketitle

\section{Introduction}
Although the standard model (SM) is the most successful theory, it is widely expected that there is new physics beyond it.
In particular, there is  an anomaly in $B^0_{s}\to \mu^+\mu^-$~\cite{CMS:2019bbr,LHCb:2017rmj,ATLAS:2018cur,LHCb:2021awg,LHCb:2021vsc}, 
in which  a 1.8 standard deviation discrepancy in the decay branching ratio between the experimental measurement and  SM's expectation~\cite{Beneke:2017vpq,Beneke:2019slt,Bobeth:2013tba,Bobeth:2013uxa,Hermann:2013kca} has been found. 
In this study, we  concentrate on 
the charmless two-body decays of  $B\to {\bf B}_n \overline{\bf B}_n^\prime$,
where $B= (B^+ , B^0 , B^0_s)$ and ${\bf B}_n^{(\prime)}$ represent the  octet baryons, given by
\begin{eqnarray}
 {\bf B}_n^{(\prime)}&=&\left(\begin{array}{ccc}
\frac{1}{\sqrt{6}}\Lambda+\frac{1}{\sqrt{2}}\Sigma^0 & \Sigma^+ & p\\
 \Sigma^- &\frac{1}{\sqrt{6}}\Lambda -\frac{1}{\sqrt{2}}\Sigma^0  & n\\
 \Xi^- & \Xi^0 &-\sqrt{\frac{2}{3}}\Lambda
\end{array}\right)\,.
\end{eqnarray}
 Since the baryons are spin-1/2 particles, the decays of $B\to {\bf B}_n \overline{\bf B}_n^\prime$
 are similar to the helicity suppressed modes of $B^0_{(s)}\to \mu^+ \mu^-$. 
It is interesting to investigate whether the suppressions exit in these baryonic channels or not. Furthermore, due to the anomaly in $B_{s}^0\to \mu^+\mu^-$, 
it is also interesting to see if similar anomalies  would also be observed  in $B\to {\bf B}_n \overline{\bf B}_n^\prime$.

 Recently, the experiments at LHCb have shown that~\cite{pdg,LHCb:2016nbc,LHCb:2017swz}
\begin{eqnarray}\label{experiment}
&&{\cal B} (B_s^0\to p \overline{p}) < 1.5 \times 10^{-8}\,,\nonumber\\
&&{\cal B} (B^0\to p \overline{p})  = (1.25 \pm 0.27 \pm 0.18 )\times 10^{-8}\,,\nonumber\\
&&{\cal B} (B^+\to p \overline{\Lambda})  = (2.4 ^ {+1.0}_{-0.8} \pm 0.3 ) \times 10^{-7}\,,
\end{eqnarray}
where the first and  second uncertainties for the last two modes are statistical and systematic, respectively.
 Although both $B_s^0 \to p \overline{p}$ and $B^+ \to p \overline{\Lambda}$ are described by the $\overline{b}\to \overline{s}$ transition,  ${\cal B}(B^+ \to p \overline{\Lambda})$ is found to be ten times larger than ${\cal B}(B^0_s\to p\overline{p})$.
We will demonstrate that the decay of  $B_s^0 \to p\overline{p}$ indeed shares the same suppression mechanism 
as that of $B^0_{s} \to \mu^+\mu^-$, which explains the unmeasured small decay branching ratio in Eq.~(\ref{experiment}).

On the other hand, most of the theoretical studies on $B\to {\bf B}_n \overline{\bf B}_n^\prime$ were performed before a decade ago~\cite{1,2,3,4,5,6,7,8,9,Chua:2003it}, with the results of ${\cal B}(B^0 \to p \overline{p})$ ranging from $10^{-5}$ to $10^{-7}$, which are significantly larger than the experimental value in Eq.~(\ref{experiment})~\cite{Cheng:2009xz}. The smallness of ${\cal B} (B^0\to p \overline{p})$
can be understood by combining the $SU(3)$ color symmetry and Fierz transformation as pointed out in Ref.~\cite{Cheng:2014qxa}.
 However,  at the current stage, the dynamical details of the strong interactions in the decays of $B\to {\bf B}_n \overline{\bf B}_n^\prime$
are still lacking. Nevertheless, these decays can  be analyzed with the symmetry in QCD of the strong interaction, which have been proved to be useful in Refs.~\cite{Hsiao:2014zza,Chua:2016aqy,Chua:2013zga}, in which their results are compatible with the experimental value of ${\cal B}(B^0 \to p \overline{p})$
in Eq.~(\ref{experiment}). However, Ref.~\cite{Hsiao:2014zza} and Refs.~\cite{Chua:2016aqy,Chua:2013zga}  have only  considered the factorizable and
nonfactorizable amplitudes, respectively.
A combined analysis on the factorizable and nonfactorizable parts is clearly required, which is one of  the main purpose of this work.

There are totally 6 topological diagrams for $B\to {\bf B}_n \overline{\bf B}_n^\prime$, given by FIG.~1, in which  the quark lines for the decays are presented, and   the hadronizations take place in the color regions. Note that the gluons propagators in FIGs.~1a, 1c and 1f are essential, otherwise the amplitudes would vanish.
\begin{figure}[b]
	\begin{subfigure}{.3\textwidth}
		\centering
 		\includegraphics[width=1\linewidth]{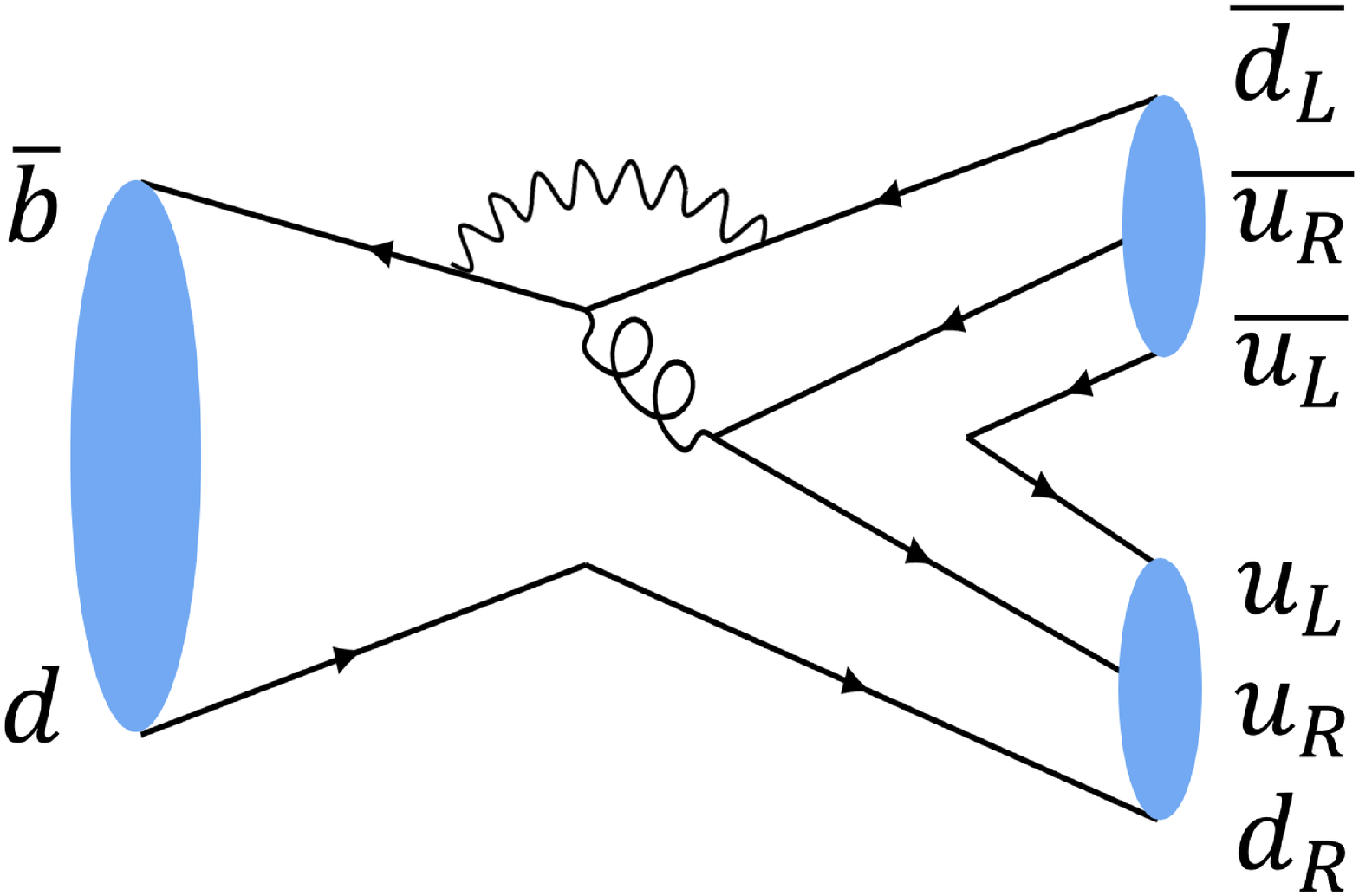}  
		\caption{ Factorizable penguin}
	\end{subfigure}
	\begin{subfigure}{.3\textwidth}
		\centering
		\includegraphics[width=1\linewidth]{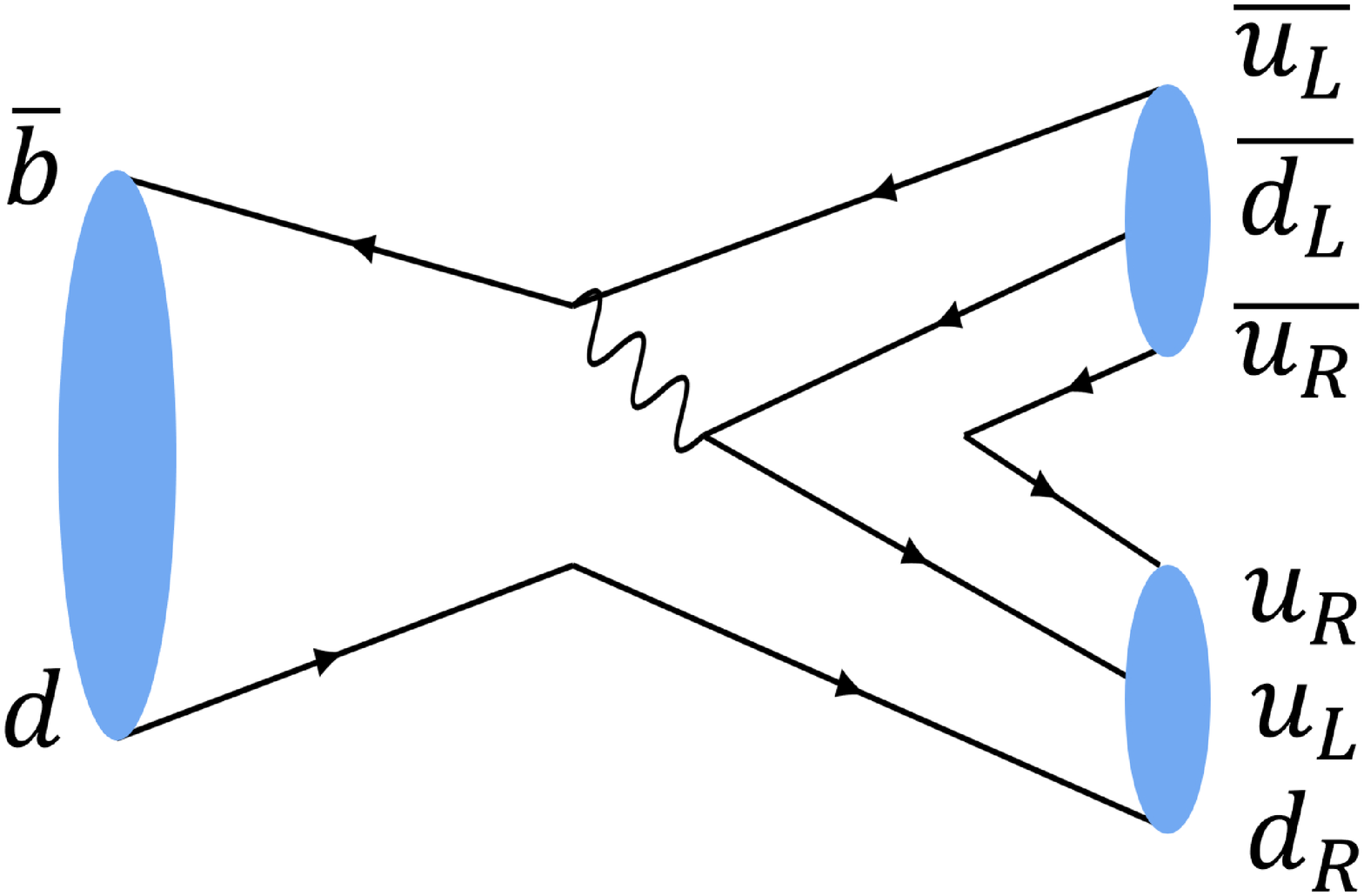}  
		\caption{Nonfactorizable tree }
	\end{subfigure}
	\begin{subfigure}{.3\textwidth}
		\centering
		\includegraphics[width=1\linewidth]{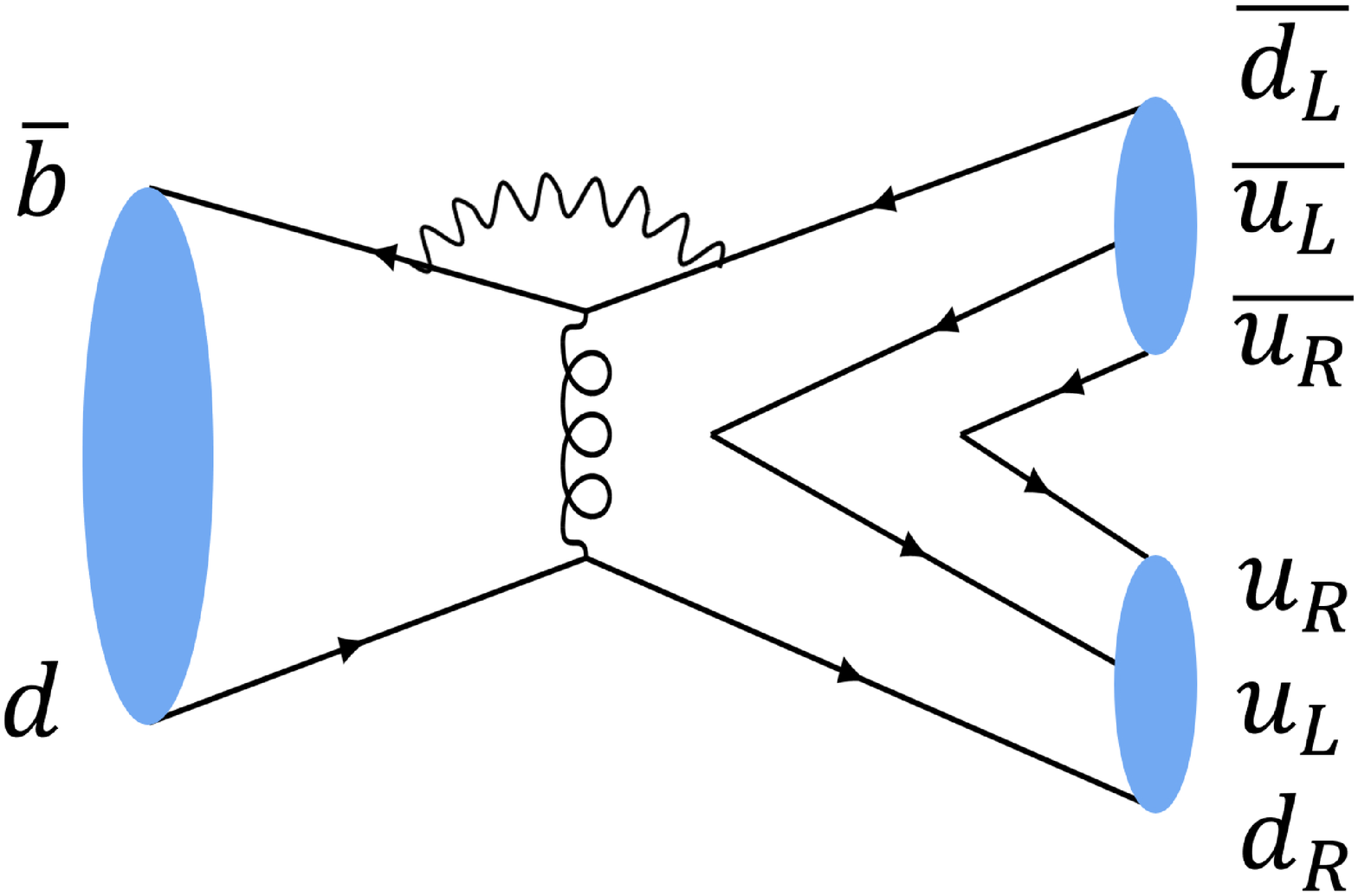}  
		\caption{Nonfactorizable penguin}
	\end{subfigure}
	\begin{subfigure}{.25\textwidth}
		\centering
		\includegraphics[width=1\linewidth]{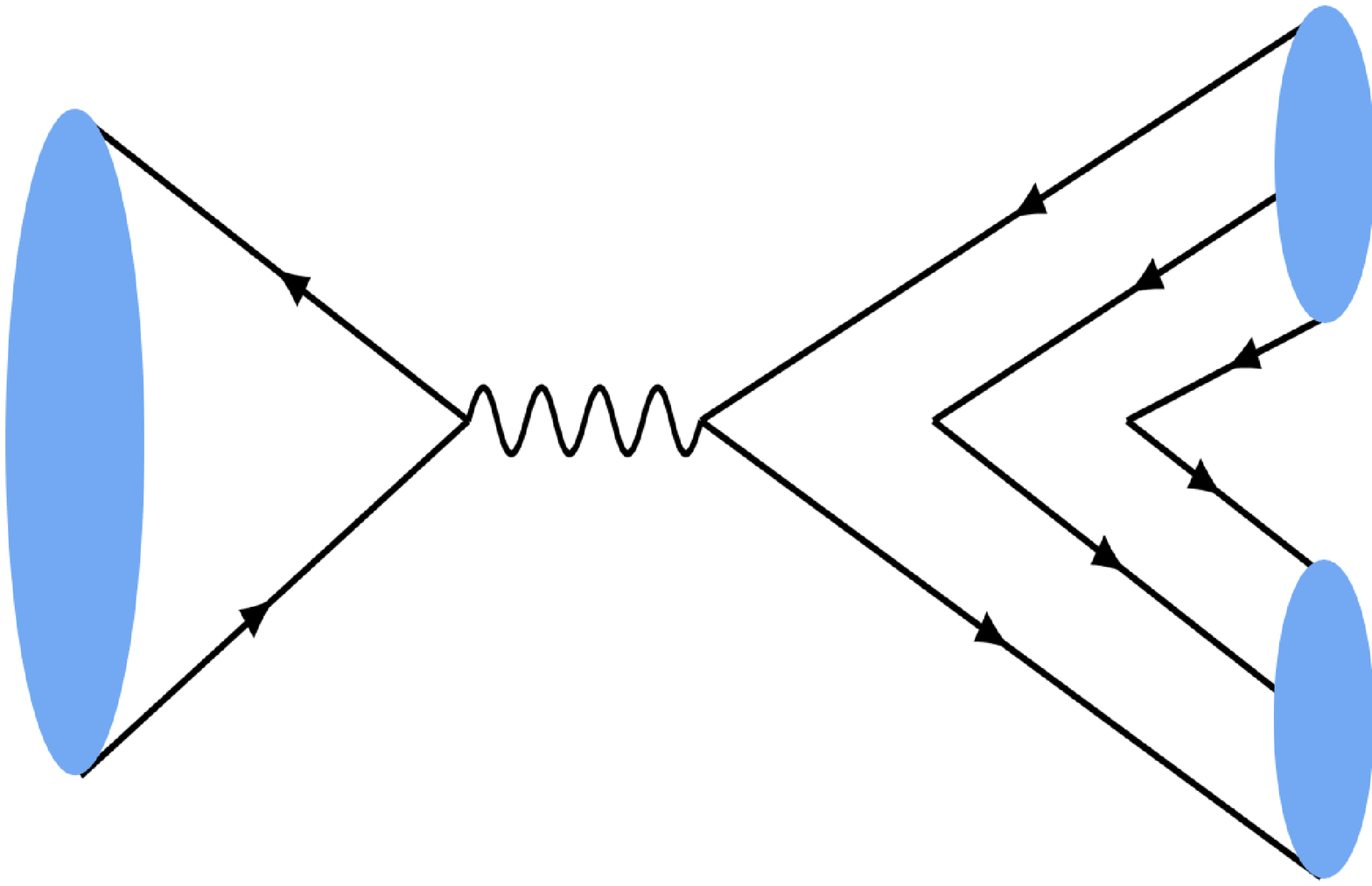}  
		\caption{Factorizable tree}
	\end{subfigure}
	~~~~~~~~~
	\begin{subfigure}{.25\textwidth}
		\centering
		\includegraphics[width=1\linewidth]{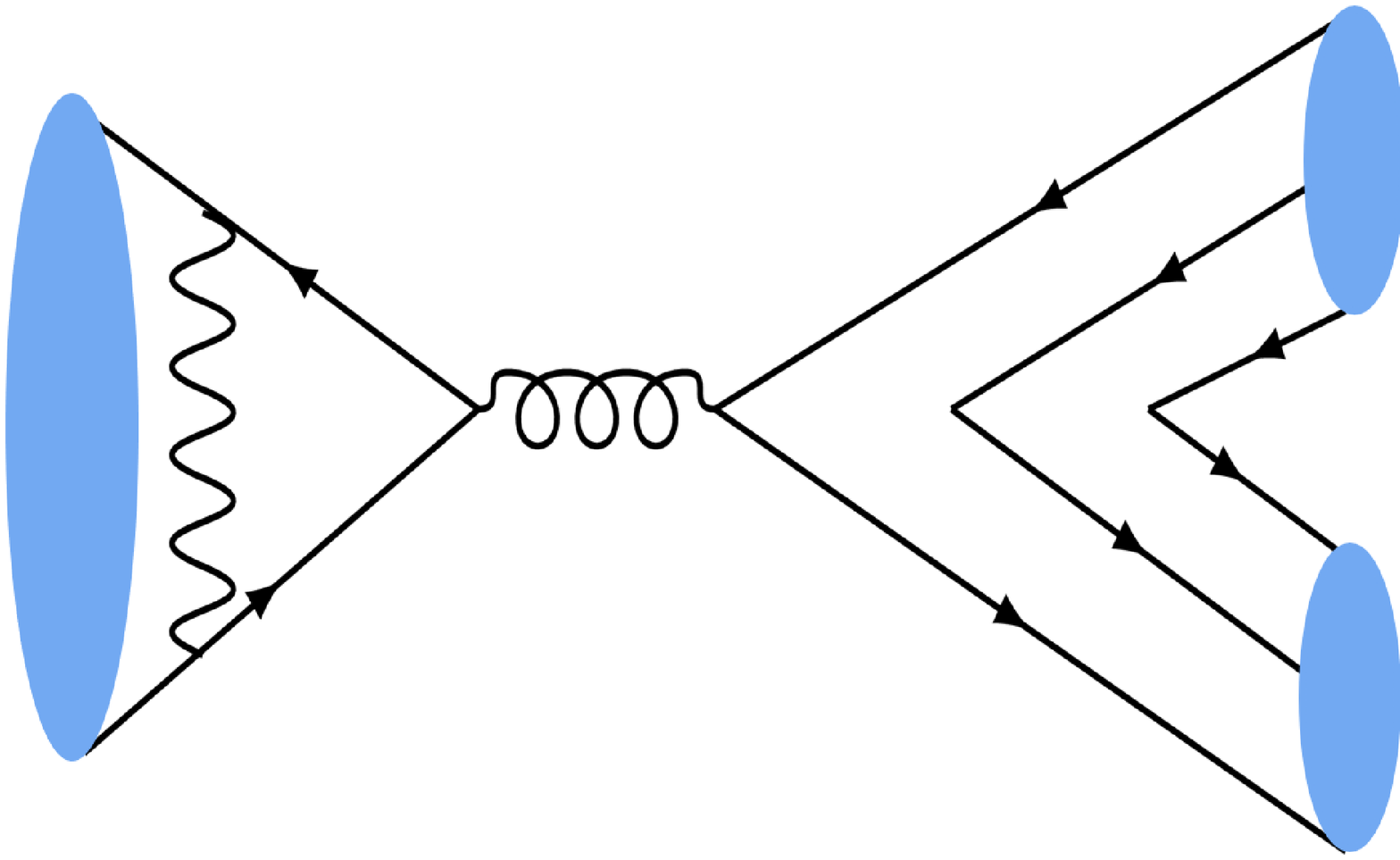}  
		\caption{Factorizable tree }
	\end{subfigure}	
	~~~~~~~~~
\begin{subfigure}{.25\textwidth}
	\centering
	\includegraphics[width=.98\linewidth]{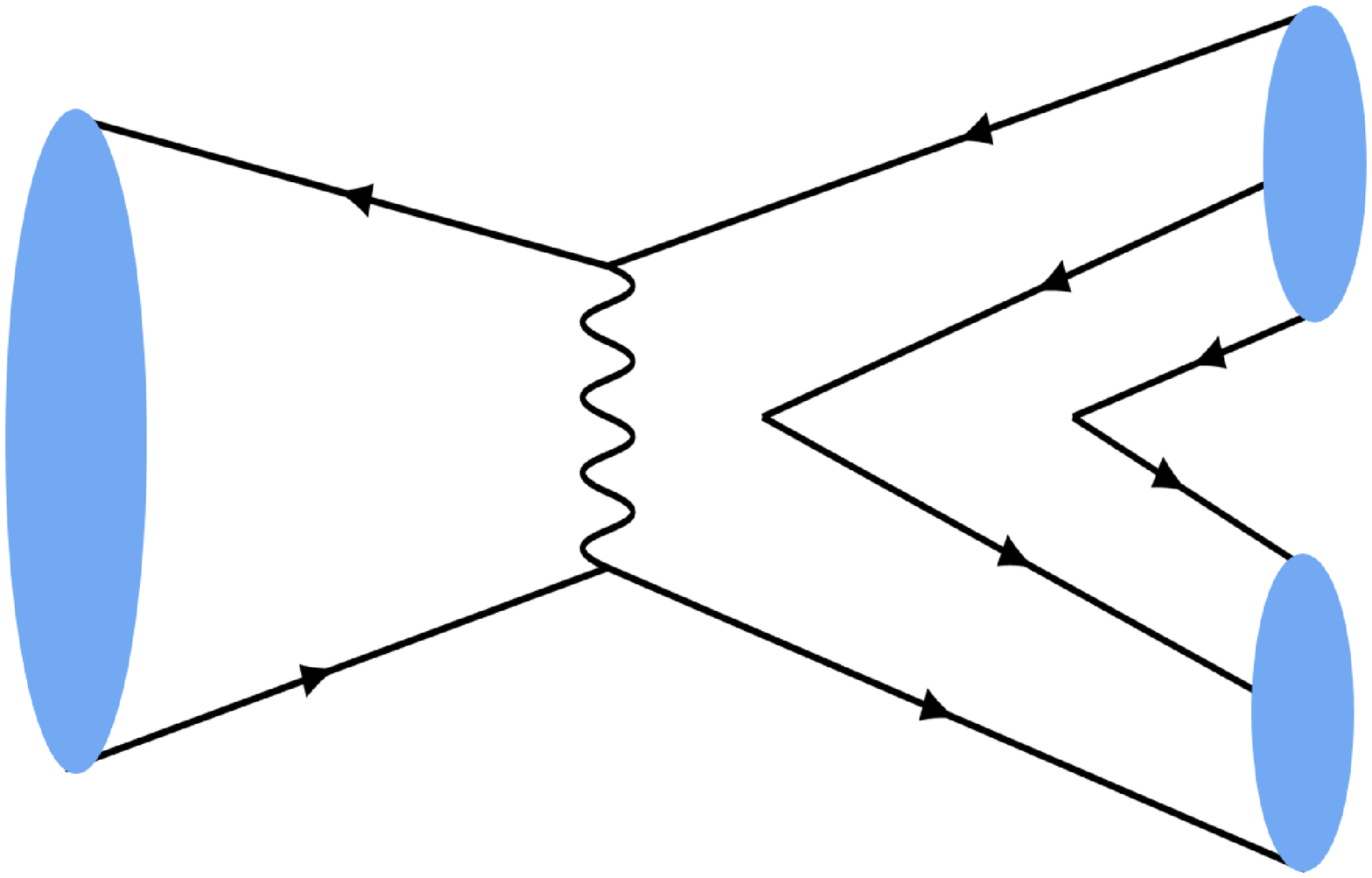}  
	\caption{Factorizable penguin }
\end{subfigure}
	\centering 
	\caption{
	Possible quark line diagrams for $B\to {\bf B}_n \overline{\bf B}_n^\prime$.}
\end{figure}
Since ${\bf B}_n$ and $\overline{\bf B}_n^\prime$ have the same helicities due to the angular momentum conservation, we  have
\begin{equation}\label{number}
N_{q_{L,R}}= N_{\overline{q}_{R,L}},
\end{equation}
in the heavy quark limit,
where $N_{q_{L,R}(\overline{q}_{R,L})}$ correspond to the numbers of $q_{L,R}~(\overline{q}_{R,L})$ in the (anti)baryons, respectively. 
The possible spin flavor configurations for $B^0\to p \overline{p}$ are shown explicitly in FIGs~1a, 1b and 1c.
In contrast,
 FIGs. 1d, 1e and 1f are not able to fulfill Eq.~\eqref{number} and therefore suppressed by the chiral symmetry~\cite{Chua:2003it}. 
 One important observation is that FIGs.~1a, 1b and 1c can only be positive helicity.
 Note that the decay of $B_s^0 \to p \overline{p}$ only receives contributions from FIGs.~1e and 1f and is suppressed consequently,
 in which the suppression mechanism is similar to $B_{(s)}^0\to \mu^+\mu^-$.

This paper is organized as follows. In Sec.~\MakeUppercase{\romannumeral 2}, we present the formalism. Our numerical results are shown in Sec.~\MakeUppercase{\romannumeral 3}. We conclude the study in Sec.~\MakeUppercase{\romannumeral 4}.

\section{formalism}

\subsection{Factorizable amplitudes}
For $\mu \ll M_W$ with $\mu $ ($M_W$) the energy scale  ($W$-boson mass), the charmless weak transition can be described by the effective Hamiltonian, given by~\cite{Buras:1991jm}
\begin{eqnarray}
&&{\cal H}_{eff} = \frac{G_F}{\sqrt{2}}
\sum_{f=d,s}\left[
V_{ub}V_{uf}^* \left(
C_1O^f_1 + C_2 O^f_2
\right) - V_{tb}V_{tq}^* \sum_{i=3}^{10}C_i O^f_i
\right]\,+ \, h.c.
\end{eqnarray}
where $G_f$ is the Fermi constant, $V_{ij}$ represent the CKM matrix elements,  $q= (u,d,s)$,
  $C_{1\text{-}10}$ correspond to the Wilson coefficients, and $O_k^f$ are the operators, defined by
\begin{equation}
\begin{array}{ll}
O_1^f =  (\overline{u}_\alpha \gamma^\mu L  b_\alpha) (\overline{f}_\beta \gamma_\mu L u_\beta) \,,\,\,\,\,&O_2^f =(\overline{u}_\alpha \gamma^\mu L b_\beta) (\overline{f}_\beta \gamma_\mu L u_\alpha)\,,\\ 
O_3^f = (\overline{f}_\alpha \gamma^\mu L b_\alpha) \sum_{q} (\overline{q}_\beta  \gamma_\mu L q_\beta)\,,\,\,\,\,&O_4^f = (\overline{f}_\alpha \gamma^\mu L b_\beta) \sum_{q}(\overline{q}_\beta \gamma_\mu L q_\alpha)\,,\\
O_5^f = (\overline{f}_\alpha \gamma^\mu L b_\alpha) \sum_{q}(\overline{q}_\beta \gamma_\mu R q_\beta)\,,\,\,\,\,&O_6^f = (\overline{f}_\alpha \gamma^\mu L  b_\beta) \sum_{q}(\overline{q}_\beta \gamma_\mu R q_\alpha)\,,\\
O_7^f = \frac{3}{2}(\overline{f}_\alpha \gamma^\mu L b_\alpha) \sum_{q}e_{q}(\overline{q}_\beta \gamma_\mu R q_\beta)\,,\,\,\,\,&O_8^f =\frac{3}{2} (\overline{f}_\alpha  \gamma^\mu L  b_\beta) \sum_{q}e_{q}(\overline{q}_\beta \gamma_\mu R q_\alpha)\,,\\
O_9^f = \frac{3}{2}(\overline{f}_\alpha  \gamma^\mu L b_\alpha) \sum_{q}e_{q} (\overline{q}_\beta \gamma_\mu L  q_\beta)\,,\,\,\,\,&O_{10}^f =\frac{3}{2} (\overline{f}_\alpha \gamma^\mu L  b_\beta) \sum_{q}e_{q}(\overline{q}_\beta \gamma_\mu L  q_\alpha)\,,
\end{array}
\end{equation}
with $f=( d,s)$, $L(R)=1\mp\gamma_5$, and the subscripts of $\alpha$ and $\beta$ being  the color indexes.
Here,  $O_{1,2}$ are the tree-level operators, while $O_{3\text{-}6}$~$(O_{7\text{-}10})$ are the (electro-)penguin ones. 
The amplitudes are then given by sandwiching ${\cal H}_{eff}$ with the initial and final states.

As $C_{7\text{-}10}$ are suppressed due to the smallness of the fine structure constant~($\alpha\approx1/137$), we would ignore  their contributions in this work.
Under the factorization framework, the amplitudes of $B\to {\bf B}_n \overline{\bf B}_n^\prime$ can be  split into two parts, described by the annihilations of the $B$ mesons and the productions of ${\bf B}_n \overline{\bf B}_n^\prime$ from the vacuum. Symbolically, they are  given by
\begin{equation}
\frac{G_F}{\sqrt{2}}{\cal C}
\langle  {\bf B}_n \overline{\bf B}_n^\prime| \overline{q}\Gamma_{\bf B} q'| 0 \rangle \langle 0 | \overline{b} \Gamma_B  q_B | B\rangle \,,
\end{equation}
where $q_B=u,d,s$  for $B= B^+, B^0, B_s^0$, respectively,  
 ${\cal C}$ are related to the CKM matrix elements and  Wilson coefficients, and $\Gamma_{{\bf B},B}$ correspond to the Dirac gamma matrices. 
 Accordingly, it is straightforward to see that FIGs.~1a, 1d, 1e and 1f  are factorizable, in which the $B$ mesons are annihilated by  $O_{1\text{-}6}$. 
 Explicitly,
 the amplitudes for FIG.~1a are given as 
\begin{eqnarray}\label{Adef}
&&A_a = \frac{G_F}{\sqrt{2}}
V_{tb}^* V_{tf} f_B
\left(
c_6+ c_5/3
\right)\frac{2m_B^2}{m_b}\langle {\bf B}_n \overline{\bf B}_n^\prime | \overline{f}R q_B |0\rangle \,,
\end{eqnarray}
where $m_{B(b)}$ are the  masses of  $B(b)$,  and $f_B$ are the $B$ meson decay constants.
Note that, due to the the charge conservation, only FIG.~1d  contributes to
the  $B^+$ decays, given by
\begin{eqnarray}
&&A_d = \frac{G_F}{\sqrt{2}}V_{ub} V_{uf}^* f_{B^+}  (c_1+c_2/3)\left[  m_u\langle {\bf B}_n \overline{\bf B}_n^\prime|  \overline{f}R u | 0\rangle - m_f\langle {\bf B}_n \overline{\bf B}_n^\prime|  \overline{f} L u | 0\rangle
\right] \,,
\end{eqnarray}
while the $B^0_{(s)}$ decays receive both contributions from FIGs.~1e and 1f, given by
\begin{eqnarray}
&&A_e = \frac{G_F}{\sqrt{2}}V_{ub} V_{uf}^* f_{B^0_{(s)}}(c_2 + c_1/3)2m_u\langle {\bf B}_n \overline{\bf B}_n^\prime|  \overline{u} \gamma_5  u |0\rangle\,,\nonumber\\
&&A_f=-\frac{G_F}{\sqrt{2}}V_{tb} V_{tf}^* f_{B^0_{(s)}} [(c_3-c_5) + (c_4 - c_6)/3]\sum_q 2 m_q \langle {\bf B}_n \overline{\bf B}_n^\prime| \overline{q}\gamma_5 q|0\rangle \,.
\end{eqnarray}
Here, to simplify the formalism,
 we have used  the quark mass hierarchies of $m_b \gg  m_{s,u,d}$
and the equations of motions,
\begin{eqnarray}\label{eqm}
&&q^\mu \langle {\bf B}_n \overline{\bf B}_n^\prime| \overline{q} \gamma_\mu q '|0 \rangle =(m_{q'}-m_q)  \langle {\bf B}_n \overline{\bf B}_n^\prime| \overline{q}  q' |0 \rangle\,,\nonumber\\
&&q^\mu \langle {\bf B}_n \overline{\bf B}_n^\prime| \overline{q} \gamma_\mu\gamma_5 q' |0 \rangle =(-m_q-m_{\overline{q}'})  \langle {\bf B}_n \overline{\bf B}_n^\prime| \overline{q}' \gamma_5 q |0 \rangle\,
\end{eqnarray}
 and
\begin{eqnarray}
&&\langle 0 | \overline{q}\gamma_5 b | B\rangle = \frac{-i m_B^2}{m_b+m_q} f_B\,,
\end{eqnarray}
with $q^\mu = p_{{\bf B}_n} ^ \mu + p_{\overline{\bf B}_n^\prime}^ \mu$.
By comparing the amplitudes, we see that $A_{d,e,f}/ A_a= {\cal O}(m_q/m_b)$, which vanish in the heavy quark limit. 
This result in  the factorization approach is consistent with that   in the  chiral analysis, {\it i.e.} $A_{d,e,f}$ can be neglected compared with $A_a$.

In the factorization framework, the perturbative effects are taken into 
account by the Wilson coefficients, while the non-perturbative ones are absorbed into $f_B$ and 
$\langle {\bf B}_n \overline{\bf B}_n^\prime | \overline{q}' \Gamma_{\bf B} q | 0 \rangle$.  In this work, the decay constants of $f_B$ are given 
from 
the evaluations in the  lattice QCD~\cite{Hughes:2017spc}. 
In addition, the baryon productions can be  simplified by the merit of the crossing symmetry, given as 
\begin{equation}
\langle {\bf B}_n ( p^\mu_{{\bf B}_n}, S_{{\bf B}_n} ) \overline{\bf B}_n^\prime ( p^\mu _{\overline{\bf B}_n^\prime}, S_{\overline{\bf B}_n^\prime}) | \overline{q} \Gamma_{\bf B} q' | 0 \rangle = \langle {\bf B}_n( p^\mu_{{\bf B}_n}, S_{{\bf B}_n} )  | \overline{q} \Gamma_{\bf B} q' |{\bf B}_n^\prime( -p^\mu_{\overline{\bf B}_n^\prime} , -S_{\overline{\bf B}_n^\prime})\rangle \,,
\end{equation}
where $S$ stands for the spin. Notice that
to bring $\overline{\bf B}_n^\prime$ from the left-handed states to the right-handed ones, we have to take the charge conjugated and flip both the 4-momenta and spins,
 resulting in that
${\bf B}_n^\prime$  are unphysical because of the negative energies.

For the ${\bf B}_n^\prime \to {\bf B}_n$ transitions, the scalar and pseudoscalar matrix elements can be parametrized by the form factors, given by 
\begin{equation}
\label{FFs}
\begin{array}{c}
\langle {\bf B}_n | \overline{q} q'| {\bf B}_n^\prime\rangle = \overline{u}_{{\bf B}_n}  f_s(q^2)u_{{\bf B}_n^\prime}\,,\\
\langle {\bf B}_n | \overline{q}\gamma_5 q'| {\bf B}_n^\prime \rangle = \overline{u}_{{\bf B}_n}  g_a(q^2) \gamma_5 u_{{\bf B}_n^\prime}\,,
\end{array}
\end{equation}
where $u$ and $\bar{u}$ represent the Dirac spinors, and $f_s(q^2)$  and $g_a(q^2)$ are the scalar and pseudoscalar form factors, respectively. 
From the inequality 
for the on-shell baryons,
\begin{equation}
\left(p_{{\bf B}_n} ^ \mu - p_{{\bf B}_n^\prime} ^ \mu\right)^2 = q^2  \le (m_{{\bf B}_n} - m_{ \overline{\bf B}_n})^2\,,
\end{equation}
it is clear that the form factors are unphysical at $q^2 = m_B^2$.

\subsection{Nonfactorizable amplitudes}

The nonfactorizable tree diagram for $B^0 \to p \overline{p}$ is given in FIG.~1b, in which the helicities of $p$ and $\overline{p}$ are found to be positive  at the chiral limit.
As a result, the quark compositions of the baryon and anti-baryon can be written  as $|d_\uparrow u _\downarrow u_\uparrow\rangle$ and $|\overline{u}_\uparrow \overline{d}_ \uparrow \overline{u} _\downarrow\rangle$, respectively,
where the arrows denote the signs of the helicities.
Then,  the  positive helicity  amplitude can be parametrized as 
\begin{equation}\label{Ab}
A_b(B^0 \to p \overline{p}) = \frac{G_f}{\sqrt{2}} V_{ub}^* V_{ud}C_T^{B^0 \to p \overline{p}} T(B_0 \to d_\uparrow u _\downarrow u_\uparrow, \overline{u}_\uparrow \overline{d}_ \uparrow \overline{u}_ \downarrow),
\end{equation}
with 
\begin{equation}\label{CT}
C_T^{B^0 \to p \overline{p}} = 
6 \langle p _\uparrow |d_\uparrow u_ \downarrow u_\uparrow\rangle  6   \langle \overline{p } _\uparrow |\overline{u}_\uparrow \overline{d}_ \uparrow \overline{u} _\downarrow\rangle\,,
\end{equation}
where $T$ stands for the tree contribution to the amplitude at the quark-level in FIG. 1b,
 $C_T$ is  the overlapping coefficient between the quarks and baryons, and $|p_\uparrow\rangle$ denotes the proton wave function given in Eq.~\eqref{proton}.
In Eq.~\eqref{CT},  the two factors of ``6'' come from that the low-lying octet baryons are symmetric in the flavor and spin indices.

To relate $A_b(B^0 \to p\overline{p})$ with
the others, we utilize the $SU(3)_F$ flavor symmetry.
It can be done systematically by the following rules with $C_T$ being modified accordingly:
\begin{itemize}
	\item $T$ would remain the same if we replace the quark line of $u$ at the bottom of FIG. 1b with either $d$ or $s$.
	\item After substituting $B^+$ or $B_s^0$ for $B^0$, $T$ is unaltered by modifying the  quark constitute of $q_B$ in B.
	\item For $A_b(\overline{b} \to \overline{s})$, we replace $\overline{d}_\uparrow (V_{ud})$ by $\overline{s}_\uparrow (V_{us})$ in 
	the right-handed side of Eq.~\eqref{Ab}.
\end{itemize}
In all, for the quark-level amplitudes, we have  
\begin{equation}\label{equality}
T(B_0 \to d_\uparrow u _\downarrow u_\uparrow, \overline{u}_\uparrow \overline{d}_ \uparrow \overline{u}_ \downarrow)=T(B \to q_{B\uparrow} u_\downarrow q_\uparrow, \overline{u}_\uparrow \overline{f}_\uparrow \overline{q}_\downarrow).
\end{equation}
Consequently, $A(\overline{b} \to \overline{f})$ can be  generally  parameterized as 
\begin{equation}
A_b(B \to {\bf B}_n \overline{\bf B}_n) = \frac{G_f}{\sqrt{2}} V_{ub}^* V_{uf}C_T^{B \to {\bf B}_n \overline{\bf B}_n} T,
\end{equation}  
with 
\begin{equation}
C_T^{B \to {\bf B}_n \overline{\bf B}_n} =36 \sum_{q}  
\langle {\bf B}_{n \uparrow} |q_{B\uparrow} u_ \downarrow q_\uparrow\rangle  \langle \overline{\bf B } _{n\uparrow }|\overline{u}_\uparrow \overline{f}_ \uparrow \overline{q} _\downarrow\rangle\,.
\end{equation}
Similarly, for $A_c(\overline{b}\to \overline{f})$, we have
\begin{eqnarray}
&&A_c(B\to {\bf B}_n \overline{\bf B}_n) = -\frac{G_f}{\sqrt{2}} V_{tb}^* V_{tf}
C_P^{B\to {\bf B}_n \overline{\bf B}_n} P 
\end{eqnarray}
with
\begin{equation}
C_P^{B\to {\bf B}_n \overline{\bf B}_n} = 36 \sum_{q,q'} \langle {\bf B}_{n\uparrow} |q_{B\uparrow} q_{ \downarrow} q'_{\uparrow}\rangle  \langle\overline{\bf B } _{n\uparrow }|\overline{f}_\uparrow \overline{q}_ {\uparrow} \overline{q}' _{\downarrow}\rangle\,,
\end{equation}
where $P$ represents the penguin contribution of FIG.~1c.

\section{Numerical results}
	To calculate the form factors  in Eq.~(\ref{FFs}), the baryon wave functions are needed.
In the  framework of the bag model, the quarks are assumed to be confined in a static bag within which the quarks move freely due to the asymptotic freedom. The only free parameter is the bag radius, which can be determined by the mass spectra. Due to its simplicity, it has been widely applied to the  baryon decays.
However, a static bag is not invariant under the space transition, and hence it could not be an eigenstate of  the 4-momentum. Such drawback makes the form factors can only be calculated at $\vec{q} = 0$. 
On the other hand,  the  modified QCD bag model shown in Ref.~\cite{Geng:2020ofy},
has tackled the problem with the liner superposition of infinite static bags, resulting in the form factors can be uniquely determined even at $\vec{q}\neq 0$. 
In particular, the experimental decay branching ratios of $\Lambda_b \to p K^-/\pi^-$ and $\Lambda_b \to \Lambda \phi$  can be well explained.
For the detailed formalism of the modified bag model, along its applications, please refer to  Refs.~\cite{Geng:2020ofy,Liu:2021rvt,Geng:2021sxe}. 

 In the calculation, we use the current quark masses of $m_{u,d,s}$ given in the Particle Data Group~\cite{pdg}, and the bag radius is taken to be $R=(5.0 \pm 0 . 2)$~GeV$^{-1}$~\cite{Geng:2020ofy,Liu:2021rvt,Geng:2021sxe}.
 The uncertainties in the bag radius would be presented as error deviations in the branching ratios.
  Note that the form factors depend little on the quark mass inputs,
 and the uncertainties in $m_{u,d,s}$ do not affect the results at the precision we consider in this work.
 
To obtain the form factors in the unphysical region, we utilize the analytic continuations from the physical ones. 
To illustrate the calculation, we take the $\Lambda \to p$ transition as an example. The $q^2$-dependences are shown in FIG.~2, 
where the dots are the calculated values in the physical region, and the lines are fitted accordingly, 
which extend to the unphysical region with the $q^2$-dependences
\begin{equation}\label{dip}
f_s(q^2) = \frac{f_s(0)}{1+ \kappa_2 q^2 +\kappa_4 q^4}\,,\quad g_a(q^2)=\frac{g_a(0)}{1+ \kappa_2' q^2 +\kappa_4^{\prime} q^4}\,,
\end{equation}
where  $\kappa_{2,4}^{(\prime)}$ are the free parameters to be fitted.
The exact values of $f_s(q^2)$ and $g_a(q^2)$  at $q^2=m_{B^+}^2$ are found to be 
\begin{equation}\label{instance}
f_s(m_{B^+}^2) = (3.9\pm 0.2) \times  10^{-3}\,,\quad g_a(m_{B^+}^2) = (4.3\pm 0.2)  \times 10^{-3}\,,
\end{equation} 
respectively. 
From Eq.~\eqref{instance}, it can be shown that both ${\bf B}_n$ and $\overline{\bf B}_n^\prime$ are predominated by the positive helicity
states,  agreeing with the chiral analysis provided in the previous section.
Assuming the nonfactorizable amplitudes are ignorable, the results of the  decay branching ratios are given in Table~\MakeUppercase{\romannumeral 1},
where   ${\cal B}_a$ correspond to  the  contributions from  FIG.~1a only.
\begin{figure}[b]
			\captionsetup{justification=raggedright,
		singlelinecheck=false
	}
		\includegraphics[width=0.7\linewidth]{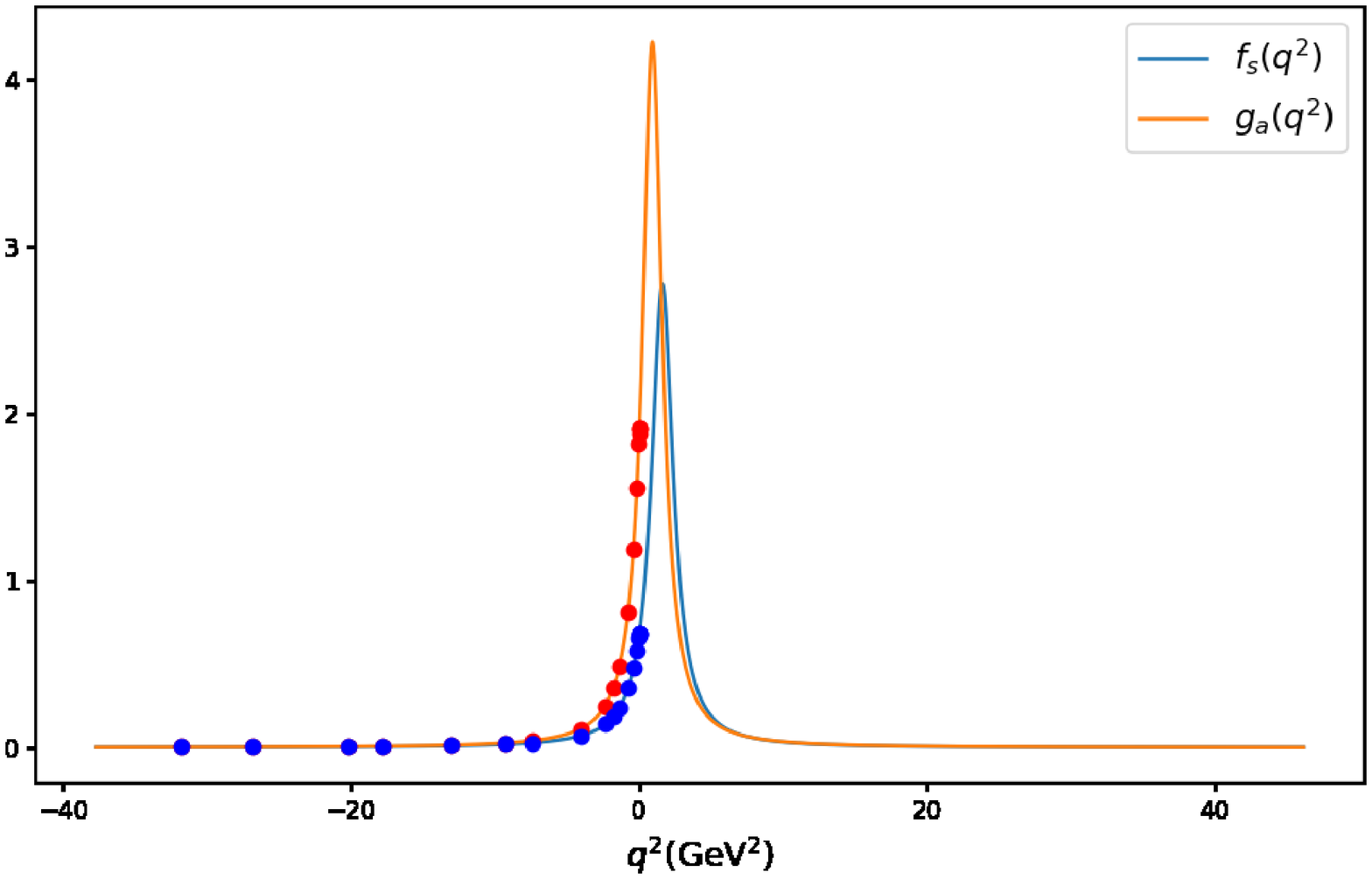}  
		\caption{ $q^2$-dependences of the form factors from the modified bag model, 
		where the results for the dots are calculated directly, while the lines are fitted by the dipole behaviors in Eq.~\eqref{dip}.}
\end{figure}
Particularly, we find that 
\begin{equation}
\begin{array}{ll}
{\cal B}_a (B^+ \to p \overline{\Lambda})  =(1.3\pm 0.1)  \times 10^{-7}\,,\quad&
{\cal B}_a (B^0 \to p \overline{p})= (2\pm 0) \times 10^{-9}\,,\\
{\cal B}_a (B^0_s \to \Xi^- \overline{\Xi}^-) =(28.8 \pm 1.5) \times 10^{-7}\,,\quad\quad
&{\cal B}_a (B^0_s \to \Xi^0 \overline{\Xi}^0) =( 27.6 \pm 1.5) \times 10^{-7}\,,\\
{\cal B}_a (B^0 \to \Lambda  \overline{\Lambda})  =(4\pm 0 )  \times 10^{-9}\,.
\end{array}
\end{equation}
Our results for the decay branching ratios on $B^+ \to p \overline{\Lambda}$ and $B^0 \to p \overline{p}$ are much smaller than the experimental values,
 showing  that the nonfactorizable effects play  leading roles,
 which justify the assumption made in Refs.~\cite{Chua:2016aqy,Chua:2013zga}.
 However, in $B_s^0\to \Xi^- \overline{\Xi}^-$ and $B_s^0\to \Xi^- \overline{\Xi}^-$, the amplitudes of $A_a$ contribute significantly, and therefore can not be neglected.
On the other hand, ${\cal B}_a( B^0 \to \Lambda \overline{\Lambda}) $ is important since it does not receive nonfactorizable contributions as we shall see  next.

Note that the matrix elements of $\langle \Sigma | \overline{q} \Gamma_{\bf B} q' |\Lambda \rangle$ vanish  with $\Sigma=( \Sigma^\pm, \Sigma^0)$, so the decays of 
$B\to (\Sigma \overline{\Lambda}, \Lambda \overline{\Sigma})$ do not receive the factorizable contributions. On the other hand, the nonfactorizable amplitudes do contribute in those decays, providing possibilities to explore  the nonfactorizable parts. However, it has to be emphasized that the argument would not hold once the $\Sigma^0- \Lambda$ mixing are considered~\cite{Geng:2020tlx},  which would cause deviations at $O(10^{-2})$ level.

\begin{table}[!h]
			\captionsetup{justification=raggedright,
	singlelinecheck=false
}
	\caption{ Results for the  decay  branching ratios, where $C_{T,P}$ are  the overlapping coefficients   and
	${\cal B}~({\cal B}_a)$ corresponds to  the (non)factorizable contributions.}
	\begin{tabular}{lccccc||lccccccc}
		\hline
 		\multicolumn{2}{c}{$\overline{b} \to \overline{s} $ }& $C_T$ & $C_P$ & $10^7 {\cal B}_a$ & $10^7 {\cal B}$ & \multicolumn{2}{c}{$\overline{b} \to \overline{d}$ } &$C_T$ & $C_P$ & $10^8 {\cal B}_a$  & $10^8 {\cal B}$ \\
 \hline
 \multirow{6}{*}{$ B^+\to$}& $ p \overline{\Sigma}^0 $&$ - \sqrt{2} $&$ \sqrt{2} $&$ 0.6 \pm 0.0 $&$  0.6 \pm 0.1 $&\multirow{6}{*}{$ B^+\to$} & $  p \overline{n} $&$ -2 $&$ -10 $&$ 0.2 \pm 0.0 $&$  1.8 \pm 0.6 $ \\
 &$p \overline{\Lambda} $&$  - \sqrt{6} $&$ - 3 \sqrt{6} $&$ 1.3 \pm 0.1 $&$  2.4 \pm 0.9 $&&$ \Sigma^+ \overline{\Sigma}^0 $&$ - 2 \sqrt{2} $&$ - 4 \sqrt{2} $&$ 2.9 \pm 0.2 $&$  4.6 \pm 0.8 $ \\
 &$n \overline{\Sigma}^- $&$ 0 $&$ -2 $&$ 1.1 \pm 0.1 $&$  0.9 \pm 0.2 $&&$\Sigma^+ \overline{\Lambda} $&$ 0 $&$ 2 \sqrt{6} $&$ 0 $&$  0.1 \pm 0.1 $ \\
 &$ \Sigma^+ \overline{\Xi}^0 $&$-2 $&$ -10 $&$ 5.0 \pm 0.3 $&$  7.7 \pm 2.3 $&&$ \Sigma^0 \overline{\Sigma}^- $&$ 0 $&$ 4 \sqrt{2} $&$ 2.9 \pm 0.2 $&$  2.2 \pm 0.5 $ \\
 &$ \Sigma^0 \overline{\Xi}^- $&$ 0 $&$ 5 \sqrt{2} $&$ 2.6 \pm 0.1 $&$  1.5 \pm 0.7 $ &&$\Xi^0 \overline{\Xi}^- $&$  0 $&$ -2 $&$ 3.5 \pm 0.2 $&$  3.3 \pm 0.3 $ \\
 &$ \Lambda \overline{\Xi}^- $&$ 0 $&$ \sqrt{6} $&$ 5.7 \pm 0.3 $&$  6.3 \pm 0.6 $ &&$\Lambda \overline{\Sigma}^- $&$ 0 $&$ 2 \sqrt{6} $&$ 0 $&$  0.1 \pm 0.1 $ \\
 
 \hline
 
 \multirow{10}{*}{$ B^0\to$}& $   p \overline{\Sigma}^+ $&$ -2 $&$ -2 $&$ 1.0 \pm 0.1 $&$  0.8 \pm 0.2$&\multirow{10}{*}{$ B^0\to$} & $ p \overline{p} $&$2 $&$ 2 $&$ 0.2 \pm 0.0 $&$  1.2 \pm 0.3 $ \\
 &$ n \overline{\Sigma}^0 $&$- 2 \sqrt{2} $&$ - \sqrt{2} $&$ 0.5 \pm 0.0 $&$  0.7 \pm 0.1  $&&$ n \overline{n} $&$ -4 $&$ -8 $&$ 0.7 \pm 0.0 $&$  5.6 \pm 1.4 $ \\
 &$ n \overline{\Lambda} $&$ - 2 \sqrt{6} $&$ - 3 \sqrt{6} $&$ 1.2 \pm 0.1 $&$ 2.4 \pm 0.9 $&&$ \Sigma^0 \overline{\Sigma}^0 $&$ -2 $&$ -4 $&$ 1.3 \pm 0.1 $&$ 2.7 \pm 0.4 $ \\
 &$ \Sigma^0 \overline{\Xi}^0 $&$ - \sqrt{2} $&$ - 5 \sqrt{2} $&$ 2.3 \pm 0.1 $&$  1.3 \pm 0.6 $&&$ \Sigma^0 \overline{\Lambda} $&$ 0 $&$ 2 \sqrt{3} $&$ 0 $&$  0.04 \pm 0.04 $ \\
 &$ \Sigma^- \overline{\Xi}^- $&$ 0 $&$ -10 $&$ 4.7 \pm 0.3 $&$  7.1 \pm 2.1$&&$\Sigma^- \overline{\Sigma}^- $&$ 0 $&$ -8 $&$ 5.3 \pm 0.3 $&$  6.6 \pm 1.1 $ \\
 &$\Lambda \overline{\Xi}^0 $&$ \sqrt{6} $&$ \sqrt{6} $&$ 5.2 \pm 0.3 $&$  5.9 \pm 0.5  $&&$ \Lambda \overline{\Sigma}^0 $&$ 2 \sqrt{3} $&$ 2 \sqrt{3} $&$ 0 $&$  3.2 \pm 0.9 $ \\
 &&&&&&&$  \Xi^- \overline{\Xi}^- $&$  0 $&$ 2 $&$ 1.6 \pm 0.0  $&$  1.3 \pm 0.1 $ \\
 &&&&&&&$ \Lambda \overline{\Lambda} $&$  0 $&$ 0 $&$ 0.4\pm 0.0  $&$ 0.4\pm 0.0 $ \\
 &&&&&&&$ \Xi^0 \overline{\Xi}^0 $&$  0 $&$ 0 $&$ 0 $&$  <10^{-3} $ \\
 &&&&&&&$ \Sigma^+ \overline{\Sigma}^+ $&$  0 $&$ 0 $&$ 0 $&$ <10^{-3} $ \\
 \hline
 
 \multirow{10}{*}{$ B_s^0 \to$}& $\Sigma^+ \overline{\Sigma}^+ $&$ 2 $&$ 2 $&$ 2.9 \pm 0.2 $&$  2.5 \pm 0.3$ &\multirow{10}{*}{$ B_s^0\to$} &$ \Sigma^+ \overline{p} $&$  -2 $&$ -2 $&$ 0.4 \pm 0.0 $&$  1.5 \pm 0.3 $ \\
 &$\Sigma^0 \overline{\Sigma}^0 $&$ 1 $&$ 2 $&$ 2.9 \pm 0.2 $&$  2.5 \pm 0.3 $&&$ \Sigma^0 \overline{n} $&$ \sqrt{2} $&$ - \sqrt{2} $&$ 0.2 \pm 0.0 $&$  0.8 \pm 0.2 $ \\
 &$ \Sigma^0 \overline{\Lambda} $&$ \sqrt{3} $&$ 0 $&$ 0 $&$   (5\pm1) 10^{-3} $&&$\Xi^0 \overline{\Sigma}^0 $&$ - 4 \sqrt{2} $&$ - 5 \sqrt{2} $&$ 1.0 \pm 0.1 $&$  9.2 \pm 2.5 $ \\
 &$\Sigma^- \overline{\Sigma}^- $&$ 0 $&$ 2 $&$ 2.9 \pm 0.2 $&$  2.5 \pm 0.3 $&&$\Xi^0 \overline{\Lambda} $&$ 0 $&$ \sqrt{6} $&$ 2.2 \pm 0.1 $&$  2.4 \pm 0.2 $ \\
 &$ \Xi^0 \overline{\Xi}^0 $&$ -4 $&$ -8 $&$ 27.6 \pm 1.5 $&$  32.5 \pm 3.7 $&&$ \Xi^- \overline{\Sigma^-} $&$ 0 $&$ -10 $&$ 2.0 \pm 0.1 $&$  3.1 \pm 0.9 $ \\
 &$\Xi^- \overline{\Xi}^- $&$ 0 $&$ -8 $&$ 28.8 \pm 1.5 $&$  33.1 \pm 3.8 $&&$  \Lambda \overline{n} $&$ - \sqrt{6} $&$ - 3 \sqrt{6} $&$ 0.5 \pm 0.0 $&$  2.5 \pm 0.6 $ \\
 &$ \Lambda \overline{\Sigma}^0 $&$ - \sqrt{3} $&$ 0 $&$ 0 $&$  (5\pm1) 10^{-3}$&& \\
 &$ \Lambda \overline{\Lambda} $&$ -3 $&$ -6 $&$ 1.6 \pm 0.1 $&$  2.6 \pm 0.8 $ \\ 
 &$ p \overline{p} $& $0 $&$ 0 $&$ 0 $& $ <10^{-3} $ &&\\ 
 &$ n \overline{n} $&$ 0 $&$ 0 $&$ 0 $& $ <10^{-3} $ &&\\ 
 \hline
\end{tabular}
\end{table}

Since the nonfactorizable effects in $T$ and $P$  could not be
reliably calculated, we may extract their values with the measured branching ratios in Eq.~\eqref{experiment}.
Explicitly, we obtain that 
\begin{equation}\label{fit}
T = \left( 6.4 \pm 1.1 \right)10^{-3}\text{GeV}^3\,,~~~~~~~~P = \left( 1.7 \pm 0.2\right)10^{-4}\text{GeV}^3,
\end{equation}
where the uncertainties come from the experimental input. 
To reduce the number of the free parameters,
 we  approximate the strong phases in $T$ and $P$ to be zero. 
A direct consequence is that the  CP violating effects are not expected, since the interference from the strong phases is lacking.
It is a good approximation for  the decays associated with the  $\overline{b}\to \overline{d}$ transition,  since these decays are dominated by the tree diagram,
whereas  the relative phase would not be important. 
On the other hand, for those with the $\overline{b}\to \overline{s}$ transition, by imposing  reasonable  bounds on the strong phases, the errors on the decay
branching ratios are bounded by $1/3$ as shown in  Appendix B, which is an acceptable range for the branching ratios as the detail of the QCD dynamic is not well understood yet. 
However,  the resulting CP asymmetries would be huge. 
 For the further discussions on the effects of the strong phases and the direct CP asymmetries, please  refer to Ref.~\cite{Chua:2016aqy}.
From Eq.~\eqref{fit}, the predicted decay  branching ratios are given in  Table~\MakeUppercase{\romannumeral 1}, labeled as ${\cal B}$. 

By taking account of the CKM matrix elements, we find that the decays associated with  $\overline{b}\to \overline{s}$ and $\overline{b}\to \overline{d}$ are  predominated by $A_c$ and $A_b$, respectively, which are consistent with the results in Ref.~\cite{Chua:2016aqy}. 
Accordingly, for the $B^+$ decays with the $\overline{b}\to \overline{d}$ transition, only the final states of $p\overline{n}$ and $\Sigma^+\overline{\Sigma}^0$ receive $A_b$~($C_T\neq0$),  and are systematically larger than the others, being good candidates to be measured in the future. 
Similarly,
$B^0 \to n \overline{n}$ and $B^0_s\to \Xi^0 \overline{\Sigma^0}$ have the largest values of $C_T$, resulting in the largest decay branching ratios,  
 in the $B^0$ and $B^0_s$ decays, respectively.

	Likewise,
	the sizable decay branching ratios with  the $\overline{b} \to \overline{s}$ transitions,
	dominated by the penguin diagrams,	
	are found to be
	\begin{eqnarray}
	&&{\cal B}( B^+ \to \Sigma^+\overline{\Xi^0})= (7.7 \pm 2.3) \times 10^{-7}\,,
	\quad {\cal B}( B^0 \to \Sigma^-\overline{\Xi^-})= (7.1 \pm 2.1) \times 10^{-7}\,,\nonumber\\
&& {\cal B} ( B_s^0 \to \Xi^0 \overline{\Xi}^0) = (32.5\pm 3.7)\times 10^{-7}\,,\quad
{\cal B} ( B_s^0 \to \Xi^- \overline{\Xi}^-) = (33.1\pm 3.8)\times 10^{-7}\,.
\end{eqnarray}
 In addition, these large values are mainly attributed by $A_a$, which can be calculated without the experimental input, making them 
 more promising to be observed in the future experiments.
On the other hand,
${\cal B}(B^0_s \to \Sigma^0 \overline{\Lambda}, \overline{\Sigma}^0 \Lambda) $
are found to be suppressed since they do not receive contributions from the penguin diagrams. 

It is interesting to see that 
the decays of  $B^0 \to ( \Xi^0 \overline{\Xi}^0, \Sigma^+\overline{\Sigma}^+)$ and $B^0_s\to (p\overline{p}, n\overline{n})$ are totally factorizable and suffer the helicity suppressions. The suppression mechanism is the same as the one in $B_{(s)}^0 \to \mu ^+ \mu^-$ as emphasized early, and a sizable decay width in the future experiments would be a signal of new physics.

 \begin{table}[b]
	\setlength\tabcolsep{4pt}
				\captionsetup{justification=raggedright,
		singlelinecheck=false
	}
	\caption{Our results comparing with those  the literature, the data are taken from the Particle Data Group~\cite{pdg},  
	while Refs.~\cite{Hsiao:2014zza} and \cite{Chua:2016aqy} only consider the facotrizable and nonfactorizable amplitudes, respectively. }
	\begin{tabular}{l|cccc}
		\hline
		$10^8{\cal B}(B\to {\bf B}_n \overline{\bf B}_n^\prime)$& data~\cite{pdg,LHCb:2016nbc,LHCb:2017swz}& ours& Ref.~\cite{Hsiao:2014zza}& Ref.~\cite{Chua:2016aqy}   \\
		\hline
		$B^0 \to p \overline{p} $ & $1.25\pm 0.27\pm 0.18$&$1.2 \pm 0.3  $& $1.4\pm 0.5 $ & $ 1.47^{+0.71}_{-0.53}\,^{+0.14}_{-0}\,^{+2.07}_{-1.16} \pm 0.12$ \\
		$B^+ \to p \overline{\Lambda} $ & $24^{+10}_{-8}\pm 3$&$24 \pm 9   $& $3.5^{+0.7}_{-0.5}\ $ & $ 24^{+10.44}_{-8.54}\,^{+2.13}_{-0}\,^{+12.48}_{-9.85} \pm 0.02$ \\
		$B^0 \to \Lambda \overline{\Lambda}$ & $ <32 $&$0.4\pm 0.0 $& $0.3\pm 0.2 $ & $ 0\pm 0\pm 0^{+0.23}_{-0}\,^{+0.0005}_{-0}$ \\
		$B_s^0 \to p \overline{p}$ & $ <1.5 $&$<0.01 $& $3.0^{+1.5}_{-1.2} $ & $ 0\pm 0\pm 0^{+0.007}_{-0}$ \\
		$B^0 \to \Xi^0 \overline{\Xi}^0 $ &-&$325\pm37$&-&$24.46_{-8.71~-0~~~-12.07-0.74}^{+10.53+3.24+16.28+0.75}$\\
		$B^0 \to \Xi^- \overline{\Xi^-} $ &-&$331\pm38$&-&$22.63_{-8.05}^{+10.02} \pm 0_{-11.36-0.71}^{+15.27+0.72}$\\
		\hline
	\end{tabular}
\end{table}

In Table~\MakeUppercase{\romannumeral 2},
we select several decay channels to compare our results with the theoretical calculations in the literature~\cite{Chua:2016aqy,Hsiao:2014zza} as well as 
 the data  from the experimental measurements~\cite{pdg,LHCb:2016nbc,LHCb:2017swz}. 
Note that  Ref.~\cite{Hsiao:2014zza} is focused only on the factorizable amplitudes, whereas 
 Ref.~\cite{Chua:2016aqy}  the nonfactorizable ones.
Here, the theoretical studies have all used the experimental branching fraction of $B^0 \to p \overline{p}$ as an input and are thus well consistent with each other.  Nonetheless,  ${\cal B}(B^+ \to \overline{\Lambda} p )$  given
 in Ref.~\cite{Hsiao:2014zza}  is too small in comparison with  the data.
For ${\cal B}(B^0 \to \overline{\Lambda} \Lambda )$, our  non-zero value agrees with that in Ref.~\cite{Hsiao:2014zza}, 
but differs  from the  zero prediction in Ref.~\cite{Chua:2016aqy}.
On the other hand,
due to the factorizable contributions, 
 our results of ${\cal B}(B_s^0 \to \Xi^0\overline{\Xi}^0, \Xi^-\overline{\Xi}^-)$  are  ten times larger than the values in Ref.~\cite{Chua:2016aqy}.
In addition,  we find that the equations of motions are well consistent with the data, which is opposed to the statement for the unsuitable use of
the equations of motions in Ref.~\cite{Hsiao:2014zza}. In particular, the decay branching ratio of $B_s^0 \to p \overline{p}$  is 
found indeed suppressed by $m_{u,d}^2/m_b^2$ and  predicted to be $<10^{-10}$, agreeing  well with the experimental upper bound.

\section{Conclusions}

We have systematically analyzed the charmless two-body decays of $B\to {\bf B}_n \overline{\bf B}_n^\prime$. 
The factorizable amplitudes have been calculated by the modified bag model along  with the crossing symmetry and analytic continuation,
while the nonfactorizable contributions  have been parametrized by  the $SU(3)_F$ flavor and chiral symmetry, and   fitted with the experimental data.
Our results are compatible with the current data, and some of the predicted branching ratios can be tested in the ongoing experiments at LHCb and BELLE-II. 
Explicitly, 
we have found that the factorizable parts in $B^+ \to p \overline{\Lambda}$ and $B^0 \to p \overline{p}$  can be safely neglected,
whereas    the factorizable ones  in $B_s^0 \to ( \Xi^0 \overline{\Xi}^0, \Xi^- \overline{\Xi}^-)$ are the main contributions with
 the sizable decay branching ratios of 
$(32.5\pm 3.7, 33.1\pm 3.8)\times 10^{-7}$,
respectively, which are promising to be observed by the experiments. 
We have also shown that
the decays associated with the  $\overline{b} \to \overline{s}$ and $\overline{b} \to \overline{d}$ transitions are
 dominated by the penguin and  tree diagrams, respectively. In particular, the decays of
 $B_s^0 \to (\Lambda \overline{\Sigma}^0, \Sigma^0 \overline{\Lambda}) $ do not receive the penguin  contributions with the suppressed 
  decay branching ratios to be
$(5\pm 1)\times 10^{-10}$,
which can be regarded as  good candidates to test our approach.
In addition, we have 
 pointed out that  the decays of $B^0 \to ( \Xi^0 \overline{\Xi}^0, \Sigma^+\overline{\Sigma}^+)$ and $B^0_s\to (p\overline{p}, n\overline{n})$ are
  helicity suppressed as  $B^0_s \to \mu^+ \mu^-$. 
It is clearly interesting to see if the future measurements on these decay modes would also show some anomalous effects  as that in $B^0_s \to \mu^+ \mu^-$. 

\section*{ACKNOWLEDGMENTS}
This work was supported in part by  the Bureau of International Cooperation, Chinese Academy of Sciences.

\appendix
\section{Baryon Wave functions}
In this appendix, we give the wave functions of ${\bf B}_n$, while those of $\overline{\bf B}_n$ can be obtained by taking the charge conjugations.
\begin{eqnarray}\label{proton}
\scriptstyle|p _\uparrow \rangle = &&\scriptstyle\frac{\sqrt{2} }{6} \left(2 | d_{ \downarrow} u_{ \uparrow} u_{ \uparrow} \rangle - | d_{ \uparrow} u_{ \downarrow} u_{ \uparrow} \rangle - | d_{ \uparrow} u_{ \uparrow} u_{ \downarrow} \rangle - | u_{ \downarrow} d_{ \uparrow} u_{ \uparrow} \rangle - | u_{ \downarrow} u_{ \uparrow} d_{ \uparrow} \rangle + 2 | u_{ \uparrow} d_{ \downarrow} u_{ \uparrow} \rangle - | u_{ \uparrow} d_{ \uparrow} u_{ \downarrow} \rangle - | u_{ \uparrow} u_{ \downarrow} d_{ \uparrow} \rangle + 2 | u_{ \uparrow} u_{ \uparrow} d_{ \downarrow} \rangle\right)\nonumber\\
\scriptstyle|n_\uparrow\rangle = &&\scriptstyle\frac{\sqrt{2} }{6}\left(| d_{ \downarrow} d_{ \uparrow} u_{ \uparrow} \rangle + | d_{ \downarrow} u_{ \uparrow} d_{ \uparrow} \rangle + | d_{ \uparrow} d_{ \downarrow} u_{ \uparrow} \rangle - 2 | d_{ \uparrow} d_{ \uparrow} u_{ \downarrow} \rangle - 2 | d_{ \uparrow} u_{ \downarrow} d_{ \uparrow} \rangle + | d_{ \uparrow} u_{ \uparrow} d_{ \downarrow} \rangle - 2 | u_{ \downarrow} d_{ \uparrow} d_{ \uparrow} \rangle + | u_{ \uparrow} d_{ \downarrow} d_{ \uparrow} \rangle + | u_{ \uparrow} d_{ \uparrow} d_{ \downarrow} \rangle\right)\nonumber\\
\scriptstyle|\Sigma^+ _\uparrow \rangle = &&\scriptstyle\frac{\sqrt{2} }{6}\left(- 2 | s_{ \downarrow} u_{ \uparrow} u_{ \uparrow} \rangle + | s_{ \uparrow} u_{ \downarrow} u_{ \uparrow} \rangle + | s_{ \uparrow} u_{ \uparrow} u_{ \downarrow} \rangle + | u_{ \downarrow} s_{ \uparrow} u_{ \uparrow} \rangle + | u_{ \downarrow} u_{ \uparrow} s_{ \uparrow} \rangle - 2 | u_{ \uparrow} s_{ \downarrow} u_{ \uparrow} \rangle + | u_{ \uparrow} s_{ \uparrow} u_{ \downarrow} \rangle + | u_{ \uparrow} u_{ \downarrow} s_{ \uparrow} \rangle - 2 | u_{ \uparrow} u_{ \uparrow} s_{ \downarrow} \rangle\right)\nonumber\\
 \scriptstyle|\Sigma^0_\uparrow\rangle=&&\scriptstyle\frac{1}{6}\left(
- | d_{ \downarrow} s_{ \uparrow} u_{ \uparrow} \rangle - | d_{ \downarrow} u_{ \uparrow} s_{ \uparrow} \rangle + 2 | d_{ \uparrow} s_{ \downarrow} u_{ \uparrow} \rangle - | d_{ \uparrow} s_{ \uparrow} u_{ \downarrow} \rangle - | d_{ \uparrow} u_{ \downarrow} s_{ \uparrow} \rangle + 2 | d_{ \uparrow} u_{ \uparrow} s_{ \downarrow} \rangle + 2 | s_{ \downarrow} d_{ \uparrow} u_{ \uparrow} \rangle + 2 | s_{ \downarrow} u_{ \uparrow} d_{ \uparrow} \rangle - | s_{ \uparrow} d_{ \downarrow} u_{ \uparrow} \rangle - | s_{ \uparrow} d_{ \uparrow} u_{ \downarrow} \rangle\right. \nonumber\\ &&\scriptstyle \qquad \left.- | s_{ \uparrow} u_{ \downarrow} d_{ \uparrow} \rangle - | s_{ \uparrow} u_{ \uparrow} d_{ \downarrow} \rangle - | u_{ \downarrow} d_{ \uparrow} s_{ \uparrow} \rangle - | u_{ \downarrow} s_{ \uparrow} d_{ \uparrow} \rangle - | u_{ \uparrow} d_{ \downarrow} s_{ \uparrow} \rangle + 2 | u_{ \uparrow} d_{ \uparrow} s_{ \downarrow} \rangle + 2 | u_{ \uparrow} s_{ \downarrow} d_{ \uparrow} \rangle - | u_{ \uparrow} s_{ \uparrow} d_{ \downarrow} \rangle\right)\nonumber\\
\scriptstyle|\Sigma^- _\uparrow \rangle=&&\scriptstyle
 \frac{\sqrt{2}}{6} \left(- | d_{ \downarrow} d_{ \uparrow} s_{ \uparrow} \rangle - | d_{ \downarrow} s_{ \uparrow} d_{ \uparrow} \rangle - | d_{ \uparrow} d_{ \downarrow} s_{ \uparrow} \rangle + 2 | d_{ \uparrow} d_{ \uparrow} s_{ \downarrow} \rangle + 2 | d_{ \uparrow} s_{ \downarrow} d_{ \uparrow} \rangle - | d_{ \uparrow} s_{ \uparrow} d_{ \downarrow} \rangle + 2 | s_{ \downarrow} d_{ \uparrow} d_{ \uparrow} \rangle - | s_{ \uparrow} d_{ \downarrow} d_{ \uparrow} \rangle - | s_{ \uparrow} d_{ \uparrow} d_{ \downarrow} \rangle\right)\nonumber\\
\scriptstyle |\Xi^- _\uparrow \rangle=&&\scriptstyle
 \frac{\sqrt{2} }{6}\left(- 2 | d_{ \downarrow} s_{ \uparrow} s_{ \uparrow} \rangle + | d_{ \uparrow} s_{ \downarrow} s_{ \uparrow} \rangle + | d_{ \uparrow} s_{ \uparrow} s_{ \downarrow} \rangle + | s_{ \downarrow} d_{ \uparrow} s_{ \uparrow} \rangle + | s_{ \downarrow} s_{ \uparrow} d_{ \uparrow} \rangle - 2 | s_{ \uparrow} d_{ \downarrow} s_{ \uparrow} \rangle + | s_{ \uparrow} d_{ \uparrow} s_{ \downarrow} \rangle + | s_{ \uparrow} s_{ \downarrow} d_{ \uparrow} \rangle - 2 | s_{ \uparrow} s_{ \uparrow} d_{ \downarrow} \rangle\right)\nonumber\\
\scriptstyle|\Xi^0 _\uparrow \rangle=&&  \scriptstyle
 \frac{\sqrt{2} }{6}\left(- | s_{ \downarrow} s_{ \uparrow} u_{ \uparrow} \rangle - | s_{ \downarrow} u_{ \uparrow} s_{ \uparrow} \rangle - | s_{ \uparrow} s_{ \downarrow} u_{ \uparrow} \rangle + 2 | s_{ \uparrow} s_{ \uparrow} u_{ \downarrow} \rangle + 2 | s_{ \uparrow} u_{ \downarrow} s_{ \uparrow} \rangle - | s_{ \uparrow} u_{ \uparrow} s_{ \downarrow} \rangle + 2 | u_{ \downarrow} s_{ \uparrow} s_{ \uparrow} \rangle - | u_{ \uparrow} s_{ \downarrow} s_{ \uparrow} \rangle - | u_{ \uparrow} s_{ \uparrow} s_{ \downarrow} \rangle\right)\nonumber\\
 \scriptstyle|\Lambda_\uparrow\rangle =&&\scriptstyle \frac{1}{\sqrt{12}}\left( | d_{ \downarrow} s_{ \uparrow} u_{ \uparrow} \rangle + | d_{ \downarrow} u_{ \uparrow} s_{ \uparrow} \rangle - | d_{ \uparrow} s_{ \uparrow} u_{ \downarrow} \rangle - | d_{ \uparrow} u_{ \downarrow} s_{ \uparrow} \rangle + | s_{ \uparrow} d_{ \downarrow} u_{ \uparrow} \rangle - | s_{ \uparrow} d_{ \uparrow} u_{ \downarrow} \rangle - | s_{ \uparrow} u_{ \downarrow} d_{ \uparrow} \rangle + | s_{ \uparrow} u_{ \uparrow} d_{ \downarrow} \rangle \right.\nonumber\\
&&\scriptstyle \qquad \left.- | u_{ \downarrow} d_{ \uparrow} s_{ \uparrow} \rangle - | u_{ \downarrow} s_{ \uparrow} d_{ \uparrow} \rangle + | u_{ \uparrow} d_{ \downarrow} s_{ \uparrow} \rangle + | u_{ \uparrow} s_{ \uparrow} d_{ \downarrow} \rangle\right) 
\end{eqnarray}

\section{Uncertainties from strong phases}

 Let $A_P$ and $A_T$ be the penguin and tree  contributions  to the amplitude of a particular decay channel, respectively.
Without lost of generality, we take $A_P$ to be real~(possibly negative) and $A_T = |A_T| e^{i\phi}$ with $\pi/2\geq \phi\geq -\pi/2$ being the  phase. The effects on the branching ratios of $\phi$  can be parameterized as
\begin{equation}
\xi \equiv 1 - \frac{{\cal B}_{nc}}{{\cal B}} = 1 - \frac{\left|A_P+|A_T|\right|^2}{|A_P+A_T|^2}\,,
\end{equation}
where the subscript of ``$nc$'' denotes that $\phi$  is taken as  zero. In general, it obeys the inequality
\begin{equation}\label{B2}
\frac{2|A_PA_T|}{A_P^2 + |A_T^2|} \geq \xi \geq -\frac{2|A_PA_T|}{A_P^2 + |A_T^2|} \,,
\end{equation}
and 
\begin{equation}\label{B3}
1\geq \xi \geq \frac{\cos\phi - 1 }{1+\cos\phi}\,,
\end{equation}
where the equations hold at $\phi = \pm \pi/2$ and $A_P=|A_T|$, respectively.
In reality, we have $A_T\gg A_P$  for the decays associated with the $\overline{b} \to \overline{d}$  transition, and therefore the strong phases shall affect little in these decays. On the other hand, for those with $\overline{b} \to \overline{s}$, the actual values  of $\xi$ are bounded by $1/3$ from Eq.~\eqref{B2}, which is within the uncertainties we consider in this work.

In addition, it is reasonable to expect that the strong phases are bounded by $|\phi| \leq \pi/4$ as found in the $B\to VV$ decays with $V =\{ \rho(770), \omega(782) , K^*(892), \phi(1020)\}$~\cite{pdg,Geng:2021lrc}. In these cases,  the inequality can be further tightened to be
\begin{equation}
 \xi  \geq  2\sqrt{2} - 3   \,,
\end{equation} 
based on Eq.~\eqref{B3},

\end{document}